\documentclass[fleqn,usenatbib]{mnras}

\usepackage{txfonts}

\usepackage{graphicx}	
\usepackage{amssymb}	
\usepackage[normalem]{ulem} 

\newcommand{\Teff}{\mbox{$T_{\mathrm{eff}}$}}
\newcommand{\logg}{\mbox{$\log g$}}
\newcommand{\Line}[3]{\Ion{#1}{#2}\,#3\,\AA}

\newcommand{\Ion}[2]{#1{\,\textsc{#2}}}

\newcommand{\Rwd}{\mbox{$R_{\mathrm{wd}}$}}
\newcommand{\Mwd}{\mbox{$M_{\mathrm{wd}}$}}

\newcommand{\Msun}{\mbox{$\mathrm{M}_{\odot}$}}

\newcounter{tref}

\usepackage[usenames,dvipsnames]{xcolor}
\usepackage{soul}

\newcommand\rrst{\bgroup\markoverwith{\textcolor{violet}{\rule[0.5ex]{3pt}{1pt}}}\ULon}

\title[Unusual abundances of the white dwarf SDSS\,J1240+6710]
{SDSS\,J124043.01+671034.68: The partially burned remnant of a low-mass white dwarf that underwent thermonuclear ignition?}

\author[B.T. G\"ansicke et al.]{
Boris T. G\"ansicke$^1$\thanks{Boris.Gaensicke@warwick.ac.uk}, Detlev Koester$^2$, Roberto Raddi$^{3,4}$, Odette Toloza$^1$, S.O. Kepler$^5$\\
$^{1}$ Department of Physics, University of Warwick, Coventry CV4 7AL, UK \\
$^{2}$ Institut f\"ur Theoretische Physik und Astrophysik, University of Kiel, 24098 Kiel, Germany\\
$^{3}$ Departament de F\'isica, Universitat Polit\`ecnica de Catalunya, c/ Esteve Terrades 5, 08860 Castelldefels
Spain\\
$^{4}$ Dr. Remeis-Sternwarte Astronomical Institute \& ECAP, Friedrich Alexander Universit\"at Erlangen-N\"urnberg, Sternwartstr. 7, 96049 Bamberg, Germany \\
$^{5}$ Instituto de F\'isica, Universidade Federal do Rio Grande do Sul, 91501-900 Porto Alegre, RS, Brazil
}

\begin{document}

\date{Accepted .... Received ....; in original form }

\pagerange{\pageref{firstpage}--\pageref{lastpage}} \pubyear{2019}

\maketitle

\label{firstpage}

\begin{abstract}
The white dwarf SDSS\,J124043.01+671034.68 (SDSS\,J1240+6710) was previously found to have an oxygen-dominated atmosphere with significant traces of neon, magnesium, and silicon. A possible origin via a violent late thermal pulse or binary interactions have been suggested to explain this very unusual photospheric composition. We report the additional detection of carbon, sodium, and aluminium in far-ultraviolet and optical follow-up spectroscopy. No iron-group elements are detected, with tight upper limits on iron, cobalt and nickel, suggesting that the star underwent partial oxygen burning, but failed to ignite silicon burning. Modelling the spectral energy distribution and adopting the distance based on the \textit{Gaia} parallax, we  infer a low white dwarf mass, $M_\mathrm{wd}=0.41\pm0.05\,\mathrm{M}_\odot$. The large space velocity of SDSS\,J1240+6710, computed from the \textit{Gaia} proper motion and its radial velocity, is compatible with a Galactic rest-frame velocity of $\simeq250$\,km/s in the opposite direction with respect to the Galactic rotation, strongly supporting a binary origin of this star. We discuss the properties of SDSS\,J1240+6710 in the context of the recently identified survivors of thermonuclear supernovae, the D$^6$ and LP\,40--365 stars, and conclude that it is unlikely related to either of those two groups. We tentatively suggest that SDSS\,J1240+6710 is the partially burned remnant of a low-mass white dwarf that underwent a thermonuclear event.
\end{abstract}


\section{Introduction}
Most stars ever born in the Universe are destined to end their lives as white dwarfs~--~Earth-sized electron-degenerate remnants with typical masses of $\simeq0.6\,\mathrm{M}_\odot$, largely made up from the ashes of hydrogen and helium fusion \citep[e.g.][]{althausetal10-1}. Because of their large surface gravities, white dwarfs undergo rapid chemical stratification \citep{schatzman45-1}, and consequently their atmospheres are composed of the lightest elements left over at the end of their prior evolution: usually hydrogen, with $\simeq20$\% of white dwarfs having helium-dominated atmospheres \citep[e.g.][]{giammicheleetal12-1}. Therefore, white dwarf spectroscopy provides, normally, no insight into the composition of the ashes of the nuclear fusion reactions that powered their progenitor stars. 

A small number of exceptions to this rule are known. Carbon was the first core element  spectroscopically detected in white dwarfs with helium-dominated atmospheres \citep[e.g.][]{liebert77-1, dufouretal05-1}, which can be dredged-up by sufficiently deep convection zones \citep{koesteretal82-2, pelletieretal86-1}. Much rarer, and only discovered thanks to the vast number of spectroscopic observations of white dwarfs by the Sloan Digital Sky Survey (SDSS, \citealt{yorketal00-1}) is the detection of oxygen \citep{liebertetal03-2}. In a handful of cases, O/C ratios vastly exceeding unity are determined,  and were interpreted as direct observational evidence for the existence of ONe-core white dwarfs \citep{gaensickeetal10-1}. 

Whereas traces of carbon and oxygen can be explained by convective dredge-up of core material, more extreme cases that defy single-star evolution are the white dwarfs with carbon-dominated atmospheres identified by \citet{dufouretal08-1}, which are possibly descending from binary mergers \citep{dunlap+clemens15-1}. Recently, two new classes of runaway stars displaying peculiar atmospheres dominated by the ashes of C-, O-, and Si-burning, have  been suggested to descend from thermonuclear supernova events: the LP\,40--365 \citep{vennesetal17-1, raddietal18-2, raddietal18-1, raddietal19-1} and D$^6$ stars \citep{shenetal18-1}. Having very low surface gravities ($\log g\simeq4.5-5.5$), both classes of stars are suggested to be low-mass white dwarfs that expanded in radius by an order of magnitude following the supernova explosions they survived. Whereas the origin of these stars is still discussed \citep[e.g.][]{baueretal19-1}, the current suggestion is that the LP\,40--365 and D$^6$ stars are the partially burned accretors \citep{vennesetal17-1, raddietal18-2, raddietal18-1, raddietal19-1}, and puffed-up donors \citep{shenetal18-1}, of thermonuclear supernovae, respectively. 

Another white dwarf with a so-far unique atmospheric composition is  SDSS\,J124043.01+671034.68 (SDSS\,J1240+6710), containing primarily oxygen, with small amounts of neon and magnesium, and traces of silicon \citep{kepleretal16-1}. Here we present follow-up far-ultraviolet and optical spectroscopy of this star, which further constrains its photospheric abundances. We also analyse its kinematics, making use of the \textit{Gaia} Data Release 2 astrometry \citep[DR2;][]{gaiaetal18-1}, and discuss its possible evolutionary history.

\section{Observations}
We obtained far-ultraviolet (FUV) spectroscopy of SDSS\,J1240+6710 using the Cosmic Origin Spectrograph (\textsc{COS}, \citealt{greenetal12-1}) on-board the \textit{Hubble Space Telescope} (\textit{HST}) on 2017 January 21. Given the far northern declination of the star, we were able to make use of \textit{HST}'s continuous viewing zone to observe uninterruptedly for four consecutive spacecraft orbits, resulting in a total exposure time of 21\,425\,sec. We used the G140L grating centred at 1105\,\AA,  covering the wavelengths 1125--2280\,\AA, though the decreasing sensitivity limits the useful range $\la 2000$\,\AA. At the time of the observations, COS was using the FUV Lifetime Position~3, resulting in a spectral resolving power of $\simeq 2000$ at 1400\,\AA. We dithered the spectrum using all for FP-POS positions to minimise the effect of fixed pattern noise. The COS spectrum of SDSS\,J1240+6710 is characterised by a very large number of strong and broad absorption lines with the notable absence of Ly\,$\alpha$ absorption, which is a typical feature of canonical white dwarfs (Fig.\,\ref{f-spec_cos}).

The \textsc{time-tag} data also provides the opportunity to correct for the \Ion{O}{i} airglow contamination. We used the timefilter task from the \textsc{costools} package version 1.2.2 to exclude data taken during daylight of the four spectra. We then used tasks from the \textsc{calcos} pipeline version 3.3.5 to extract the night-side spectra and combine them into an average spectrum which excludes the airglow of O\,{\sc i}. To preserve the maximum signal-to-noise ratio of the COS observations, we only substituted in the average spectrum the region affected by O\,{\sc i} airglow with the night side data.

\begin{figure}
\centerline{\includegraphics[width=\columnwidth]{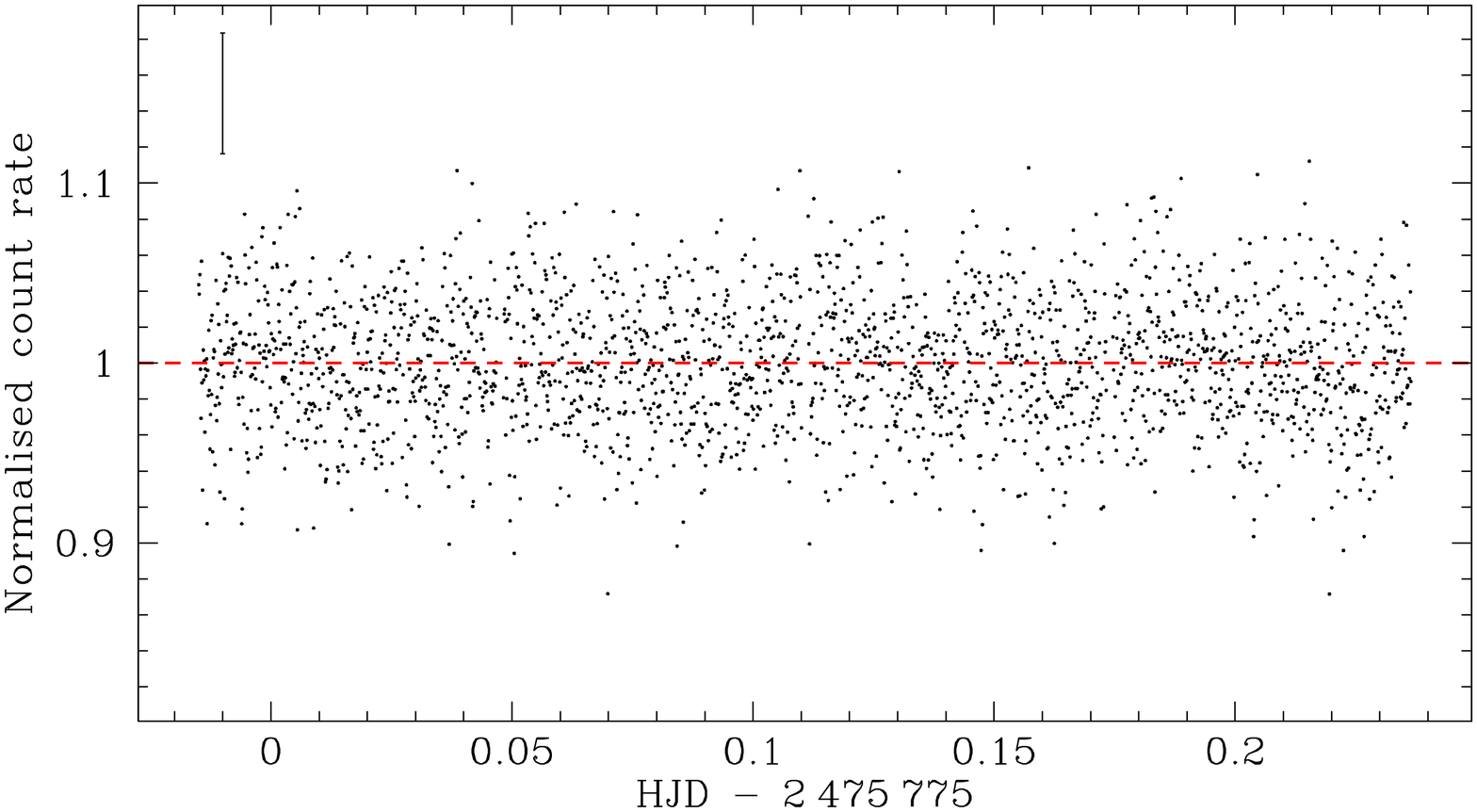}}
\caption{\label{f-lc_cos} The far-ultraviolet light curve of SDSS\,J1240+6710 constructed from the time-tagged COS observations, binned in 10\,s. Time-series analysis of these data rules out variability with amplitudes larger than $\simeq0.5$\,\% and periods of $\simeq30$\,s to 6\,h. The error bar illustrates the typical uncertainty.}
\end{figure}

The \textsc{COS} data were obtained in the \textsc{time-tag} mode, registering wavelength and arrival time of each individual photon, which allows to construct the ultraviolet light curve of SDSS\,J1240+6710. The source counts were extracted over the wavelength range 1145--1443\,\AA, using a box with a height of 51 pixel, centred on the spectral trace of the target. Airglow emission lines are present in the average COS observations (Fig.\,\ref{f-spec_cos}), and we therefore masked out the strong Lyman\,$\alpha$ and the \Ion{O}{i} airglow lines in the range of 1208.18--1223.33\,\AA\ and  1295.50--1313.06\,\AA, respectively.  The background was extracted using two regions above and below the target spectrum, both with a height of 31 pixels. The background counts, scaled to the relative areas of the extracted regions, were subtracted from the source counts. The extracted source and background regions were corrected for the time-dependent sensitivity function using the reference files associated with the observations\footnote{The reference files can be downloaded directly from \href{ftp://ftp.stsci.edu/cdbs/lref}{ftp://ftp.stsci.edu/cdbs/lref}}, converting the count rates into fluxes. Finally, the resulting background-subtracted light curve is averaged in 10\,s bins (Fig.\,\ref{f-lc_cos}).

\begin{figure}
\centerline{\includegraphics[width=\columnwidth]{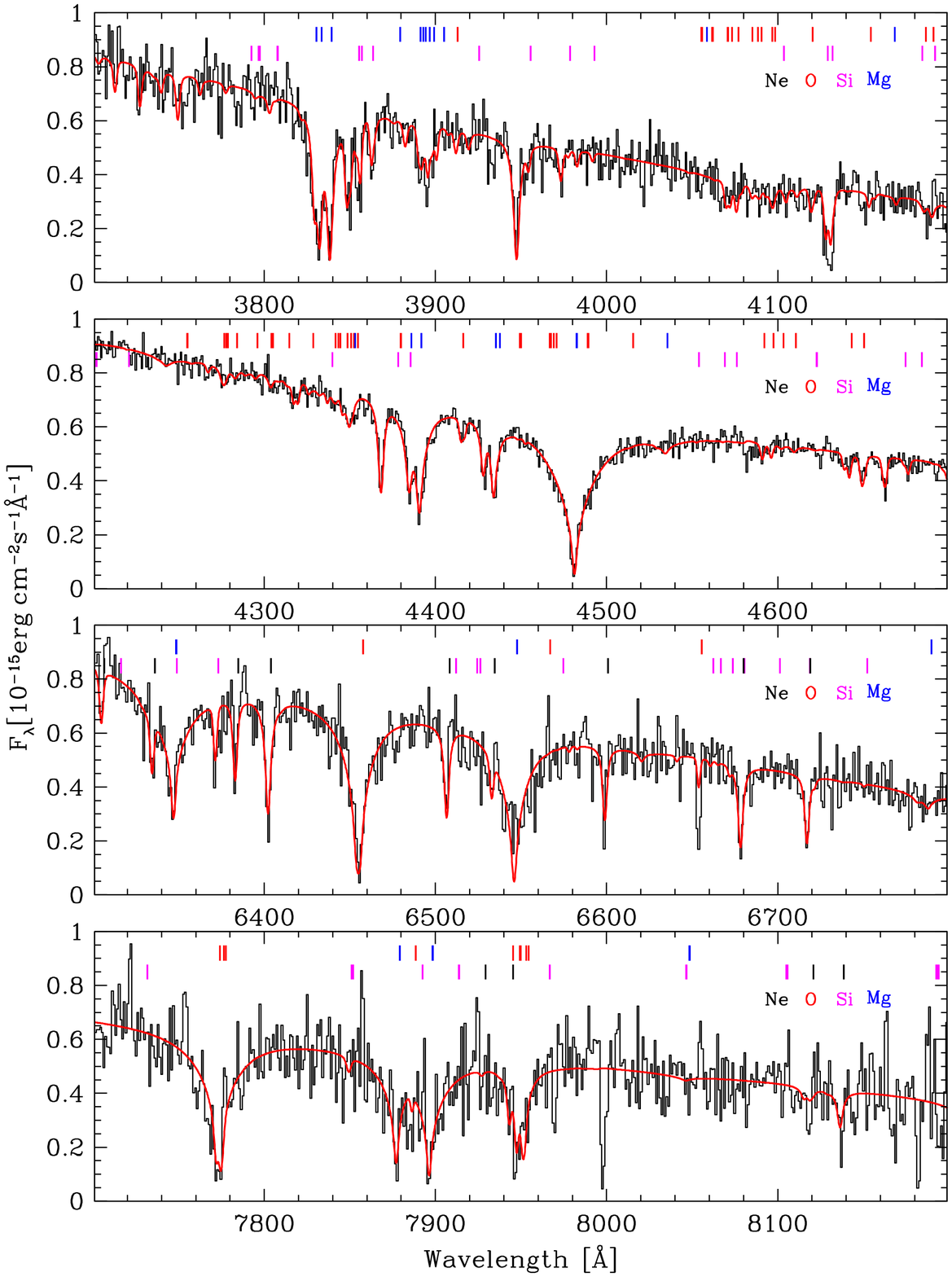}}
\caption{\label{f-spec_wht} The WHT spectrum of SDSS\,J1240+6710 (black) contains all the spectral features already present in the SDSS spectrum \citep{kepleretal16-1}. The detection of the neon lines is more robust in the new data. The best-fit white dwarf model (Tables\,\ref{t-parameters}), \ref{t-abundances} is shown in red. The absorption feature near 8000\,\AA\ results from imperfect  telluric correction.}
\end{figure}

Optical spectroscopy of SDSS\,J1240+6710 was obtained on 2017 April 2 using the double-arm Intermediate-dispersion Spectrograph and Imaging System (ISIS) on the William Herschel Telescope (WHT). We used the R600B and R600R gratings in the blue and red arm, respectively, with a one-arcsec slit. The red arm was equipped with a GG495 order-blocking filter. This set-up provides a spectral resolution of $\simeq2$\,\AA\ in both arms of the spectrograph. We obtained a total of seven 20\,min exposure pairs with central wavelengths at 3930\,\AA\ and 6561\,\AA\ in the blue and red arm, respectively, and five 20\,min exposures pairs with central wavelengths of 4540\,\AA\ and 8201\,\AA. These four wavelength settings resulted in a spectral coverage of $3270-5410$\,\AA\ and $5820-9180$\,\AA. The observations were carried out under $\simeq 1$\,arcsec seeing and good transparency, and were interleaved with arc-lamp calibrations every hour. The spectra were reduced following standard techniques and using the \textsc{pamela}\footnote{github.com/starlink/starlink/tree/master/applications/pamela} and \textsc{molly}\footnote{deneb.astro.warwick.ac.uk/phsaap/software/} packages. The WHT spectrum (Fig.\,\ref{f-spec_wht}) is of better spectral resolution, and higher signal-to-noise ratio (median of 21) than the SDSS discovery spectrum  \citep{kepleretal16-1}.

\section{Atmospheric analysis}
In addition to the new spectroscopic observations that we obtained, a key piece of information that has become available since the discovery of SDSS\,J1240+6710 is its distance, $d=439\pm30$\,pc, based on the \textit{Gaia} DR2 parallax (Table\,\ref{t-parameters}, \citealt{gaiaetal18-1}).

\begin{figure*}
\centerline{\includegraphics[width=0.8\textwidth]{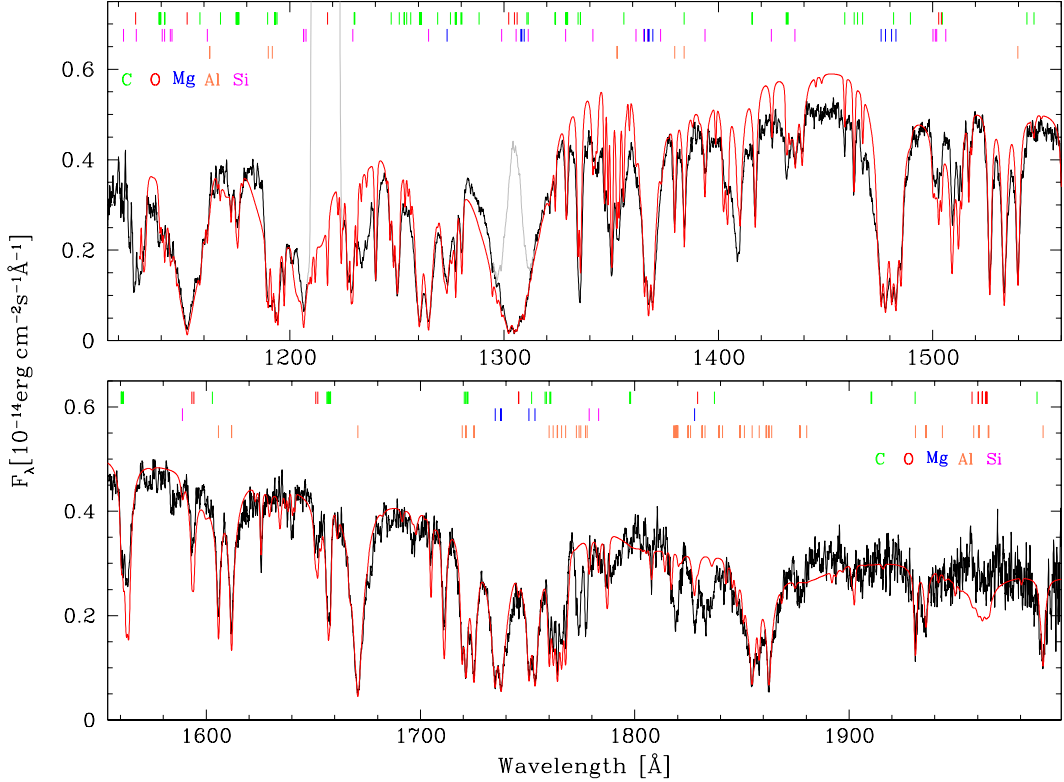}}
\caption{\label{f-spec_cos} The far-ultraviolet \textit{HST}/COS spectrum of SDSS\,J1240+6710  (black) contains many strong absorption lines of C, O, Mg, Al and Si, as indicated by the coloured tick marks. Geocoronal airglow of Ly$\alpha$ and \Line{O}{i}{1302} contaminates the spectrum (gray sections); the latter can be removed by using the data obtained on the night-side of the \textit{HST} orbit around the Earth. The flux of the best-fit white dwarf model (red) exceeds the observed spectrum in the range $\simeq1300-1500$\,\AA, which we suspect to be related to missing continuum opacities. A small number of absorption features remain unidentified near 1233, 1773, 1777, 1818, 1827, and 1833\,\AA.} 
\end{figure*}

Because of the excellent absolute flux calibration of the COS spectroscopy, we used these data to refine the atmospheric parameters of SDSS\,J1240+6710. We began the 
spectral analysis based on the atmospheric parameters determined by \citet{kepleretal16-1}, i.e. $\Teff=21\,590$\,K, $\log g=7.93$, and manually adjusted the abundances of the main elements contributing to the ultraviolet absorption spectrum (C, O, Mg, Al, Si) to obtain a reasonable match to the observed line strengths. 

We cross-correlated this initial model with the COS data to measure the radial velocity of SDSS\,J1240+6710, which we found to be $\varv_\mathrm{rad}=-158\pm12$\,km/s. We accounted for this blue-shift in all the following spectral analysis. With the starting model described above, we modelled the available spectroscopy, distance, and SDSS $ugriz$ photometry (which is given in Table\,1 of \citealt{kepleretal16-1}). This approach uses the spectral slope over a wide wavelength range by combining the optical photometry, and the absolutely flux-calibrated ultraviolet spectrum, and was carried out iteratively as follows.

(1) We computed a model with the starting $\Teff$, $\log g$, and C, O, Mg, Al, and Si abundances, applied reddening\footnote{We use the reddening as provided by SDSS, which corresponds to the total extinction along the line of sight based on \citet{schlegeletal98-1}.} corresponding to $A_g=0.0647$, and calculated synthetic $ugriz$ magnitudes. The difference between the observed $(m_\mathrm{o})$ and synthetic magnitudes ($m_\mathrm{s}$) is then
\begin{equation}
m_\mathrm{o}-m_\mathrm{s} = -2.5 \log\Omega~~~\mbox{with the solid angle}~~~\Omega=\pi\left(\frac{\Rwd}{d}\right)^2
\end{equation}

\noindent 
With the distance determined by \textit{Gaia}, the scaling factor above implies a white dwarf radius, \Rwd. Using the helium-atmosphere white dwarf mass-radius relation\footnote{For a given $\Teff$ and $\log g$, white dwarfs with a non-degenerate hydrogen layer have slightly larger radii and lower surface gravities than helium-atmosphere white dwarfs. Given that the atmosphere of SDSS\,J1240+6710 contains no significant amount of hydrogen, adopting the mass-radius relation for helium-atmosphere white dwarfs is more appropriate. However, some systematic uncertainty arises from the unknown core composition of SDSS\,J1240+6710.}
of the Montreal group\footnote{\label{fn-cooling}http://www.astro.umontreal.ca/$\sim$bergeron/CoolingModels/, \citet{holberg+bergeron06-1, kowalski+saumon06-1, tremblayetal11-2, bergeronetal11-1}.}, we calculated the mass of the star, which then gives an updated value of $\log g$. 

(2) If the difference between the initial value of $\log g$ and the new one differ by more than 0.1\,dex, the new value is adopted. 

(3) The iterated and reddened model is multiplied with the solid angle $\Omega$, and compared to the absolutely flux-calibrated \textit{HST}/COS spectrum. The flux level depends on both $\log g$ (via the radius) and on $\Teff$. As long as the predicted flux level was significantly too high (low), we went back to (1) with a lower (higher) $\Teff$ (keeping the abundances of C, O, Mg, Al, and Si fixed at their initial values). 

The modelling procedure converged for $\Teff=20\,500\pm 500$\,K, $\log g=7.62\pm 0.11$, and $\Mwd=0.41\pm 0.05\,\mathrm{M}_\odot$. The quoted uncertainties in the surface gravity and the white dwarf mass are combined from the uncertainties in both $T_\mathrm{eff}$ and the distance. We find that SDSS\,J1240+6710 has a deep convection zone below an optical depth $\tau=2/3$ enclosing a mass fraction of $\log(M_\mathrm{CVZ}/M_\mathrm{wd})\simeq-5.3$.

With $\Teff$ and $\log g$ fixed, we proceeded to refine the abundance analysis of the COS spectrum. The relative abundances of O, Mg, and Si measured from the COS data are consistent with those determined by \citet{kepleretal16-1} from the optical SDSS spectrum. Given the strength and  large number of the ultraviolet lines, our new values supersede those from the previous study due to an improved accuracy and  precision. In addition to those elements, we detect C, and Al in the COS spectrum, and Na in the optical spectrum. We also determined upper limits for an additional 15 elements from the respective strongest absorption lines predicted in the model spectrum (see Fig.\,\ref{f-abundances} and Table\,\ref{t-abundances}). 

The optical \textsc{WHT} spectrum does not provide additional constraints on the abundances of O, Mg, and Si compared to the analysis of the COS data. However, given its higher spectral resolution and better signal-to-noise ratio compared to the SDSS discovery spectrum \citep{kepleretal16-1}, we were able to refine the abundance measurement of Ne, which has no ultraviolet transitions (Table\,\ref{t-abundances}). The new Ne abundance is consistent with, but somewhat higher than that of \citet{kepleretal16-1}.

Given the highly unusual atmospheric composition of this white dwarf, we note a few caveats to our analysis. A practical problem is that while the high quality \textit{HST}/COS spectrum has a good spectral resolution, there  are so many lines, most of them unresolved, that it is practically impossible to define a continuum. It is rarely possible to fit individual lines, but the entire spectrum has to be calculated consistently, with all lines of all elements included.

More fundamental problems relate to the available atomic data. The ultraviolet and optical spectra of SDSS\,J1240+6710 contain a very large number of absorption lines of O, Mg, Al, Si, and Ne. Naturally this includes many lines that are weak or absent altogether in the spectra of normal stars. The atomic parameters, oscillator strength  and broadening constants, are therefore not well determined, or not available at all. We retrieved line lists from VALD \citep{piskunovetal95-1} and NIST \citep{kramidaetal16-1}. Comparing the available atomic data, we found that on most cases the values from the NIST database were more reliable and internally consistent, though we complemented our line list with data from VALD where no information was available on NIST. 

VALD includes a large amount of line broadening data, which is very useful for the analysis. However, those data are, as appropriate for all normal applications, calculated for hydrogen atoms as perturbers. At the temperature of SDSS\,J1240+6710 line broadening is predominantly Stark broadening. For the heavy elements this is dominated by electrons. Our usual procedure assumes electrons and an equal number of ionised hydrogen as perturbers. Since the contribution of the ions decreases with atomic weight, this should be reasonably accurate also for an oxygen, neon, magnesium-dominated atmosphere.

A final comment concerns the continuum opacities from the photo-ionisation of C, O, Ne, Mg, Al, and Si. Whereas these cross-sections are available from TOPBASE, the Opacity Project database \citep{cuntoetal93-1}, they are not as well known and tested as the hydrogen and helium opacities, e.g. the wavelengths of predicted absorption edges can differ from the detected ones by as much as $10-20$\,\AA. More importantly, the TOPBASE cross-sections include many very large resonances, which are often not observed. We follow a common approach \citep{bautistaetal98-1} and re-bin the TOPBASE cross-sections to smooth out the strongest peaks. However, there remain broad absorption features in the model without counterparts in the observations, and with $>2000$ cross-sections included in the model, it is very difficult to identify those with problematic opacity data. The mis-match between the continuum flux of our best-fit model around $\simeq1300-1500$\,\AA\ is likely caused by remaining problems in the choice of cross-sections we adopted (Fig.\,\ref{f-spec_cos}). Despite all efforts in identifying the absorption lines in the spectrum of SDSS\,J1240+6710, unaccounted features remain at 1233, 1773, 1777, 1818, 1827, 1833, 3999, 4011, and 4023\,\AA.

\begin{table}
\caption{\label{t-parameters} Main characteristics of SDSS\,J1240+6710.}
\centering
\begin{tabular}{llr@{$\,\pm\,$}l}\hline
Parameter & symbol & \multicolumn{2}{c}{value}\\
\hline
Parallax               & $\varpi$\,[mas]         & 2.2805      & 0.1573 \\
Proper motion          & $\mu_\alpha$\,[mas/yr]  & $-184.343$  & 0.293	\\
                       & $\mu_\delta$\,[mas/yr]  & $-95.776$   & 0.236 \\
Tangential velocity    & $\varv_\perp$\,[km/s]       & 432         & 30 \\
Radial velocity        & $\varv_\mathrm{rad}$\,[km/s]  & $-158$      & 12   \\
Gravitational redshift & $\varv_\mathrm{gr}$\,[km/s] & 16.5        & 1   \\
Rest-frame velocity    & $\varv_\mathrm{rest}$\,[km/s] & \multicolumn{2}{c}{$\simeq250$} \\
Distance               & $d$\,[pc]               & 439         & 30  \\
Effective temperature  & \Teff\,[K]              & 20\,500     & 500 \\
Surface gravity        & \logg\ (cgs)            & 7.62        & 0.11  \\
Mass                   & \Mwd\,[\Msun]           & 0.41        & 0.05  \\ 
Mass of convection zone& $\log(M_\mathrm{cvz}/M_\mathrm{wd})$ & \multicolumn{2}{c}{$-5.3$}\\
Cooling age            & $\tau_\mathrm{cool}$\,[Myr] & \multicolumn{2}{c}{$\sim 40$}\\ \hline
\multicolumn{3}{l}{Note: Mass, convection zone, and cooling age are derived from}\\
\multicolumn{3}{l}{He-dominated  models.}
\end{tabular}
\end{table}

\begin{table}
\caption{\label{t-abundances}
Photospheric number abundances relative to oxygen.}
\centering
\begin{tabular}{cccc}\hline
Element & $\log[\mathrm{Z/O}]$ & Element & $\log[\mathrm{Z/O}]$ \\ \hline
H  & $<-3.30$       & Ca & $<-6.00$ \\
He & $<-1.80$       & Sc & $<-5.80$ \\ 
C  & $-3.44\pm0.14$ & Ti & $<-7.50$ \\
N  & $<-5.50$       & V  & $<-5.30$ \\
Ne & $-1.15\pm0.09$ & Cr & $<-5.00$ \\
Na & $-2.30\pm0.30$ & Mn & $<-5.50$ \\
Mg & $-1.95\pm0.10$ & Fe & $<-5.50$ \\
Al & $-3.80\pm0.16$ & Co & $<-5.50$ \\
Si & $-3.43\pm0.30$ & Ni & $<-5.80$ \\ 
P  & $<-5.80$       & Cu & $<-7.80$ \\
S  & $<-4.80$       & Sr & $<-6.00$ \\
Ar & $<-3.50$ \\       
\hline
\end{tabular}
\end{table}

\section{Probing for photometric and radial velocity variability}
We used the strong and sharp photospheric absorption lines in the optical spectra of SDSS\,J1240+6710 to search for radial velocity variations. The individual WHT spectra were obtained over a baseline of four hours, and their radial velocities are consistent with each other to $\la 30$\,km/s. Comparing the average WHT spectrum with the SDSS spectrum obtained about three years earlier, we can rule out long-term radial variations in excess of $\simeq 50$\,km/s. In conclusion, the COS, WHT and SDSS spectroscopy consistently show that the star is blue-shifted by $-158\pm12$\,km/s.

Subjecting the COS far-ultraviolet light curve (Fig.\,\ref{f-lc_cos}) to a time-series analysis, we do not detect any significant periodicity, with an upper limit of a fractional amplitude $\la 0.5$\% for periods in the range of $\simeq 30$\,s to $\simeq 6$\,h. The Zwicky Transient Factory \citep{mascietal19-1} obtained 28 (34) $g-$band ($r$-band) images of SDSS\,J1240+6710 between March and December 2018, which also do not show any variability in excess of 0.05\,mag. 

We conclude that there is no evidence for SDSS\,J1240+6710 being member of a short-period binary system. 

\section{Kinematics}
The \textit{Gaia} proper motions and distance of SDSS\,J1240+6710 imply a tangential velocity of $432\pm 30$\,km/s.  Given its low mass, the gravitational redshift of the white dwarf is $\simeq 17$\,km/s. Correcting for that the true radial velocity of SDSS\,J1240+6710 becomes $-177\pm 10$\,km/s; hence, its space velocity  with respect to the Sun is $466\pm 30$\,km/s. 

We modelled the Galactic trajectory of SDSS\,J1240+6710 taking into account a standard formulation for the Milky Way's  potential, composed  by a power-law density profile with exponential cut-off for the bulge, a Miyamoto-Nagai disc, and a dark matter halo  (implemented as \verb|MWPotential2014| in the {\sc python} module {\sc galpy}; \citealt{bovy15-1}). The assumed Galactic potential implies an escape velocity of $\simeq 560$\,km/s in the Solar neighbourhood (in agreement recent estimates from {\em Gaia} DR2; \citealt{monarietal18-1}). We adopt a Galactic non-rotating rest-frame that is left-handed, with the $x$-axis pointing from the Galactic centre to the Sun, and the $y$-axis pointing towards the direction of the Galactic rotation. The Sun is at $8.27$\,kpc from the centre.

We have simulated 10\,000 trajectories of SDSS\,J1240+6710 by sampling the {\em Gaia} astrometry and the radial-velocity as priors for $t = 0$ (i.e. now) and taking into account their mutual correlations. Assuming a Galactic rotation of $239\pm 9$\,km/s \citep{schoenrich12-1} and canonical values for the Solar motion components \citep{schoenrichetal10-1}, our simulation shows that SDSS\,J1240+6710 is moving around  the Milky Way against the average rotation. This white dwarf is following an ellipsoidal trajectory in the Galactic rest-frame ($e \simeq 0.55$) that ranges between $z = \pm 0.8$\,kpc and $R_{G} =4$--$13$\,kpc, which are the vertical and radial coordinates, respectively, in the Galactocentric  cylindrical reference frame. The rest-frame velocity of SDSS\,J1240+6710 is $\simeq 250$\,km/s and its vertical component is  $V_z = -15 \pm 11$\,km/s.
Taking about  240\,Myr to complete a full revolution around the Galactic centre, SDSS\,J1240+6710 may have been on such an unusual orbit for a relatively short time of the order of its estimated cooling age ($\approx 40$\,Myr).

We note that the retrograde  orbit of SDSS\,J1240+6710 is reminiscent of the motion of halo white dwarfs that, however, on average have much larger amplitudes in their vertical ($z$) displacements (see \citealt{paulietal06-1}).

\begin{figure}
\centerline{\includegraphics[width=\columnwidth]{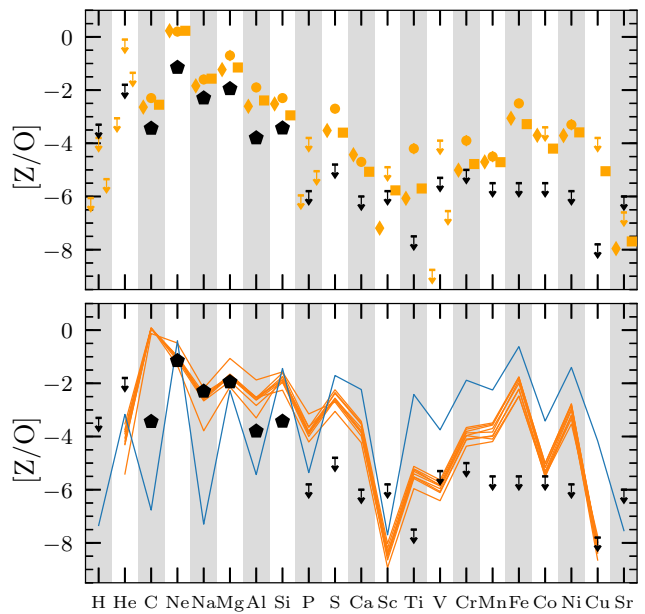}}
\caption{\label{f-abundances} Top panel: comparison of the photospheric number abundances of SDSS\,J1240+6710 relative to oxygen (black) and the three LP\,40--365 stars (orange) from \citet{raddietal19-1}. Bottom panel: the bulk-composition of bound remnants resulting from three dimensional hydrodynamic simulations of pure deflagrations of CO white dwarfs (orange, \citealt{finketal14-1}) and ONe white dwarfs (blue, \citealt{jonesetal19-1}). The most noticeable characteristic of SDSS\,J1240+6710 is the non-detection of iron-group elements, of which large abundances are predicted by the models, and detected in the LP\,40--365 stars.}
\end{figure}

\section{Discussion}

\subsection{Comparison to the LP\,40--365 and D$^6$ stars}
The  relatively high space velocity of SDSS\,J1240+6710, combined with its unusual photospheric abundances and anti-clockwise Galactic orbit, are at least superficially, reminiscent of two new types of white dwarfs that were recently discovered: the LP\,40--365  stars \citep{vennesetal17-1, raddietal18-1, raddietal18-2, raddietal19-1} and the D$^6$ stars \citep{shenetal18-1}. 

The LP\,40--365  stars all have very low masses, $\simeq$0.2--0.3\,$\mathrm{M}_\odot$, extended radii of  0.2--0.6\,$\mathrm{R}_{\odot}$, temperatures of $\simeq 10\,000$--13\,000\,K, and space velocities of $\simeq 500-650$\,km/s. The fact that they all have ONeMg-dominated atmospheres, sprinkled with other $\alpha$ and iron-peak elements (see Fig.\,\ref{f-abundances}; \citealt{raddietal19-1}) indicates that these stars are likely the partially burnt remnants of white dwarfs in close binaries that survived peculiar thermonuclear supernovae \citep{foleyetal13-1}, even though alternative explanations have been suggested \citep{baueretal19-1}. While somewhat less well characterised at the moment, the D$^6$ stars share some properties of the LP\,40--365  stars, i.e. they are significantly over-luminous compared to canonical white dwarfs, suggesting that they also have very low masses while having temperatures of $\simeq 8000$\,K. While no detailed abundance measurements are yet available, just gauging from their spectra, the D$^6$ stars are also extremely rich in photospheric transitions of many elements, including calcium \citep{shenetal18-1}. However, their space velocities are much higher than those of the LP\,40--365  stars, $\ga 1000$\,km/s, and \cite{shenetal18-1} proposed that the D$^6$ stars are the surviving donor stars of thermonuclear supernovae, unbound when the binaries got disrupted at the moment of the explosion. 

\citet{kepleretal16-1} estimated the surface gravity of SDSS\,J1240+6710 from the model atmosphere analysis of the SDSS spectrum, $\log g=7.93\pm 0.17$, which together with a mass-radius relation implied a mass of $0.56\pm0.09\mathrm{M}_\odot$~--~clearly too low for an ONe-core white dwarf, but well within the range of field white dwarfs~--~and discussed the possible origin within single-star evolution scenarios. However, our new mass determination of $0.41\pm 0.05\,\mathrm{M}_\odot$ is too low for any canonical single-star evolution model. This evidence, combined with the high space velocity, retrograde Galactic orbit, and unusual photospheric abundances, suggests that a binary origin involving some kind of thermonuclear event.

SDSS\,J1240+6710 differs from the LP\,40--365  and D$^6$ stars  in many aspects, having a higher mass, being hotter, and most strikingly, having an abundance pattern dominated by $\alpha$-elements, but no detection of iron-group elements, with tight upper limits on Ti, Fe, and Ni (Fig.\,\ref{f-abundances}). A hypothetical explanation for the lack of iron-group elements is diffusion, with iron diffusing out of the envelope faster than other elements. We crudely estimate the diffusion time scales at the bottom of the deep convection zone to be $\sim10$\,Myr  \citep{koester09-1}\footnote{Updated tables are available at http://www1.astrophysik.uni-kiel.de/$\sim$koester/astrophysics/astrophysics.html}. Assuming that SDSS\,J1240+6710 underwent a thermonuclear event that resulted in significant re-heating (which is also postulated for a non-igniting  merger event, see  \citealt{wegg+phinney12-1} and \citealt{temminketal19-1}), we can crudely estimate the time since this event by adopting a standard cooling track for a $\Teff=20\,500$\,K and $0.41\,\mathrm{M}_\odot$ white dwarf, resulting in $\sim40$\,Myr~--~which is comparable to the diffusion time scales at the bottom of the convection zone. We conclude that the photospheric abundances, reflecting the composition of the deep convective layer in the outer envelope, have probably not been altered significantly by the differential diffusion velocities of the individual elements. We also note that invoking diffusion to eliminate iron from the photosphere would also affect the relative abundances of the detected lighter elements, which do not show any striking anomaly when compared to the LP\,40--365 stars (Fig.\,\ref{f-abundances}). It appears, hence, that  SDSS\,J1240+6710 underwent nuclear burning up to $^{16}$O+$^{16}$O, producing Si, but did not proceed to silicon burning. 

\subsection{A thermonuclear event of a low-mass white dwarf?}

The thermonuclear ignition of CO white dwarfs as well as the resulting nuclear yields  have been studied in excruciating detail because of their link to supernovae type~Ia  (\citealt{nomotoetal84-1, woosleyetal86-1}, see the reviews by \citealt{nomoto+leung17-2, seitenzahl+townsley17-1})\footnote{The outcome of electron capture in a ONe white dwarf has been studied from a theoretical point of view, and initial models suggested that this pathway will predominantly result in an accretion induced collapse \citep{nomoto+condo91-1}. However, the three-dimensional oxygen deflagration models of \citet{jonesetal16-1} showed instead of a collapse an incomplete thermonuclear explosion can ensue, ejecting $\simeq1\,\Mwd$ and leaving behind a bound remnant. More recent calculations confirm these outcomes \citep{jonesetal19-1, schwab+rocha19-1}, and suggest that these remnants should be ONeFe white dwarfs. We include the nuclear yields from \citet{jonesetal19-1} in Fig.\,\ref{f-abundances}, but given the even larger predicted abundances of iron-group elements rule out that SDSS\,J1240+6710 is the remnant of a thermonuclear event involving a ONe white dwarf}. Whereas thermonuclear supernovae can be produced by growing a white dwarf to the Chandrasekhar limit \citep{nomoto+leung17-1}, most current models assume that the detonation of a helium shell triggers the ignition of the underlying CO core (the double-detonation model,  \citealt{shenetal18-2}), which can be  less massive than the Chandrasekhar limit. 

Unsurprisingly, most theoretical efforts have concentrated on those configurations that produce sufficient amounts of $^{56}$Ni to match the observed peak luminosities and decline times of SN\,Ia. \cite{polinetal19-1} systematically explored the nuclear yields of double-detonations of CO white dwarfs with masses from 0.6 to 1.2\,$\mathrm{M}_\odot$ and helium shell masses from 0.01 to 0.1\,$\mathrm{M}_\odot$, and found that for core masses $\simeq0.8\,\mathrm{M}_\odot$ only very small amounts of nickel are synthesised. Such thermonuclear events would probably be difficult to detect because of the lack of a radioactive decay powered light curve following the short initial flash or shock breakout. 

Assuming that SDSS\,J1240+6710 originates from a thermonuclear supernovae in a binary star, it must have either lost sufficient mass during the explosion to unbind the binary, or the donor star has been disrupted by the supernova ejecta. As a simple example, a 0.8\,\Msun\ CO white dwarf accreting from a 0.2\,\Msun\ He white dwarf would undergo mass transfer at a period of $\simeq5$\,min. Accretion and ignition of $\sim0.1$\,\Msun\ with the subsequent ejection of $\sim0.5$\,\Msun\ would successfully disrupt the binary.  The orbital velocity of the CO white dwarf at the moment of the explosion would be $\simeq280$\,km/s, compatible with the rest-frame velocity of SDSS\,J1240+6710.

\section{Conclusions}
We have obtained high-quality ultraviolet and optical spectroscopy of the white dwarf SDSS\,J1240+6710, and, combined with the \textit{Gaia}~DR2 proper motions and parallax of the star, determined that it has a low mass, $\Mwd=0.41\pm 0.05\,\mathrm{M}_\odot$. The star has rest-frame velocity of $\simeq250$\,km/s, but a substantially higher space velocity, $466\pm 30$\,km/s, as it rotates against the average Galactic disc motion. Its oxygen-dominated atmosphere is rich in $\alpha$-elements, and we reported improved abundance measurements of Si, Mg, and Ne as well as the additional detection of C, Na, and Al. We do not detect any iron-group element, with tight limits on the abundances of Ti, Fe, and Ni, and conclude that the star underwent oxygen burning, but did not reach the ignition conditions for silicon burning. The low mass of the white dwarf and its moderately high rest-frame velocity suggest an origin involving a thermonuclear supernova in a compact binary. The lack of iron-group elements in its atmosphere clearly distinguishes SDSS\,J1240+6710 from the two other recently discovered classes of supernova survivors, and suggests that it may be the result of the thermonuclear ignition of a low-mass, $\la0.8$\,\Msun, white dwarf. The very low mass of Ni produced and ejected in such events would make their detection extremely challenging within the current time-domain surveys.

\section*{Acknowledgements}
We are grateful to Evan Bauer for discussions regarding the possible progenitor of SDSS\,J1240+6710, and Abigail Polin for sharing unpublished material. Based on observations made with the NASA/ESA \textit{Hubble Space Telescope}, obtained at the Space Telescope Science Institute, which is operated by the Association of Universities for Research in Astronomy, Inc., under NASA contract NAS 5-26555. These observations are associated with program 14600. Based on observations made with the William Herschel Telescope operated on the island of La Palma by the Isaac Newton Group of Telescopes in the Spanish Observatorio del Roque de los Muchachos of the Instituto de Astrof\'isica de Canarias. OT was supported by a Leverhulme Trust Research Project Grant. BTG and OT were supported by the UK STFC grant ST/P000495 and ST/T000406/1. RR has received funding from the postdoctoral fellowship programme Beatriu de Pin\'os, funded by the Secretary of Universities and Research (Government of Catalonia) and by the Horizon 2020 programme of research and innovation of the European Union under the Maria Sk\l{}odowska-Curie grant agreement No 801370. RR acknowledges funding from the German Science Foundation (DFG) through grants HE1356/71-1 and IR190/1-1.

\bibliographystyle{mnras}
\bibliography{aamnem99,aabib,proceedings}

\begin{thebibliography}{}
\makeatletter
\relax
\def\mn@urlcharsother{\let\do\@makeother \do\$\do\&\do\#\do\^\do\_\do\%\do\~}
\def\mn@doi{\begingroup\mn@urlcharsother \@ifnextchar [ {\mn@doi@}
  {\mn@doi@[]}}
\def\mn@doi@[#1]#2{\def\@tempa{#1}\ifx\@tempa\@empty \href
  {http://dx.doi.org/#2} {doi:#2}\else \href {http://dx.doi.org/#2} {#1}\fi
  \endgroup}
\def\mn@eprint#1#2{\mn@eprint@#1:#2::\@nil}
\def\mn@eprint@arXiv#1{\href {http://arxiv.org/abs/#1} {{\tt arXiv:#1}}}
\def\mn@eprint@dblp#1{\href {http://dblp.uni-trier.de/rec/bibtex/#1.xml}
  {dblp:#1}}
\def\mn@eprint@#1:#2:#3:#4\@nil{\def\@tempa {#1}\def\@tempb {#2}\def\@tempc
  {#3}\ifx \@tempc \@empty \let \@tempc \@tempb \let \@tempb \@tempa \fi \ifx
  \@tempb \@empty \def\@tempb {arXiv}\fi \@ifundefined
  {mn@eprint@\@tempb}{\@tempb:\@tempc}{\expandafter \expandafter \csname
  mn@eprint@\@tempb\endcsname \expandafter{\@tempc}}}

\bibitem[\protect\citeauthoryear{{Althaus}, {C{\'o}rsico}, {Isern}  \&
  {Garc{\'{\i}}a-Berro}}{{Althaus} et~al.}{2010}]{althausetal10-1}
{Althaus} L.~G.,  {C{\'o}rsico} A.~H.,  {Isern} J.,   {Garc{\'{\i}}a-Berro} E.,
   2010, \mn@doi [\aapr] {10.1007/s00159-010-0033-1}, \href
  {2010A&ARv..18..471A} {18, 471}

\bibitem[\protect\citeauthoryear{{Bauer}, {White}  \& {Bildsten}}{{Bauer}
  et~al.}{2019}]{baueretal19-1}
{Bauer} E.~B.,  {White} C.~J.,   {Bildsten} L.,  2019, \mn@doi [ApJ]
  {10.3847/1538-4357/ab4ea4}, \href {2019ApJ...887...68B} {887, 68}

\bibitem[\protect\citeauthoryear{{Bautista}, {Romano}  \& {Pradhan}}{{Bautista}
  et~al.}{1998}]{bautistaetal98-1}
{Bautista} M.~A.,  {Romano} P.,   {Pradhan} A.~K.,  1998, \mn@doi [ApJS]
  {10.1086/313132}, \href
  {https://ui.adsabs.harvard.edu/abs/1998ApJS..118..259B} {118, 259}

\bibitem[\protect\citeauthoryear{{Bergeron} et~al.,}{{Bergeron}
  et~al.}{2011}]{bergeronetal11-1}
{Bergeron} P.,  et~al., 2011, \mn@doi [ApJ] {10.1088/0004-637X/737/1/28}, \href
  {2011ApJ...737...28B} {737, 28}

\bibitem[\protect\citeauthoryear{{Bovy}}{{Bovy}}{2015}]{bovy15-1}
{Bovy} J.,  2015, \mn@doi [ApJS] {10.1088/0067-0049/216/2/29}, \href
  {https://ui.adsabs.harvard.edu/abs/2015ApJS..216...29B} {216, 29}

\bibitem[\protect\citeauthoryear{{Cunto}, {Mendoza}, {Ochsenbein}  \&
  {Zeippen}}{{Cunto} et~al.}{1993}]{cuntoetal93-1}
{Cunto} W.,  {Mendoza} C.,  {Ochsenbein} F.,   {Zeippen} C.~J.,  1993, A\&A,
  \href {1993A&A...275L...5C} {275, L5}

\bibitem[\protect\citeauthoryear{{Dufour}, {Bergeron}  \& {Fontaine}}{{Dufour}
  et~al.}{2005}]{dufouretal05-1}
{Dufour} P.,  {Bergeron} P.,   {Fontaine} G.,  2005, \mn@doi [ApJ]
  {10.1086/430373}, \href {2005ApJ...627..404D} {627, 404}

\bibitem[\protect\citeauthoryear{{Dufour}, {Fontaine}, {Liebert}, {Schmidt}  \&
  {Behara}}{{Dufour} et~al.}{2008}]{dufouretal08-1}
{Dufour} P.,  {Fontaine} G.,  {Liebert} J.,  {Schmidt} G.~D.,   {Behara} N.,
  2008, \mn@doi [ApJ] {10.1086/589855}, \href {2008ApJ...683..978D} {683, 978}

\bibitem[\protect\citeauthoryear{{Dunlap} \& {Clemens}}{{Dunlap} \&
  {Clemens}}{2015}]{dunlap+clemens15-1}
{Dunlap} B.~H.,  {Clemens} J.~C.,  2015, in {Dufour} P.,  {Bergeron} P.,
  {Fontaine} G.,  eds,  Astronomical Society of the Pacific Conference Series
  Vol. 493, 19th European Workshop on White Dwarfs. p.~547

\bibitem[\protect\citeauthoryear{{Fink} et~al.,}{{Fink}
  et~al.}{2014}]{finketal14-1}
{Fink} M.,  et~al., 2014, \mn@doi [MNRAS] {10.1093/mnras/stt2315}, \href
  {2014MNRAS.438.1762F} {438, 1762}

\bibitem[\protect\citeauthoryear{{Foley} et~al.,}{{Foley}
  et~al.}{2013}]{foleyetal13-1}
{Foley} R.~J.,  et~al., 2013, \mn@doi [ApJ] {10.1088/0004-637X/767/1/57}, \href
  {2013ApJ...767...57F} {767, 57}

\bibitem[\protect\citeauthoryear{{Gaia Collaboration} et~al.,}{{Gaia
  Collaboration} et~al.}{2018}]{gaiaetal18-1}
{Gaia Collaboration} et~al., 2018, \mn@doi [A\&A]
  {10.1051/0004-6361/201833051}, \href {2018A&A...616A...1G} {616, A1}

\bibitem[\protect\citeauthoryear{{G{\"a}nsicke}, {Koester}, {Girven}, {Marsh}
  \& {Steeghs}}{{G{\"a}nsicke} et~al.}{2010}]{gaensickeetal10-1}
{G{\"a}nsicke} B.~T.,  {Koester} D.,  {Girven} J.,  {Marsh} T.~R.,   {Steeghs}
  D.,  2010, \mn@doi [Science] {10.1126/science.1180228}, \href
  {2010Sci...327..188G} {327, 188}

\bibitem[\protect\citeauthoryear{{Giammichele}, {Bergeron}  \&
  {Dufour}}{{Giammichele} et~al.}{2012}]{giammicheleetal12-1}
{Giammichele} N.,  {Bergeron} P.,   {Dufour} P.,  2012, \mn@doi [ApJS]
  {10.1088/0067-0049/199/2/29}, \href {2012ApJS..199...29G} {199, 29}

\bibitem[\protect\citeauthoryear{{Green} et~al.,}{{Green}
  et~al.}{2012}]{greenetal12-1}
{Green} J.~C.,  et~al., 2012, \mn@doi [ApJ] {10.1088/0004-637X/744/1/60}, \href
  {2012ApJ...744...60G} {744, 60}

\bibitem[\protect\citeauthoryear{{Holberg} \& {Bergeron}}{{Holberg} \&
  {Bergeron}}{2006}]{holberg+bergeron06-1}
{Holberg} J.~B.,  {Bergeron} P.,  2006, \mn@doi [AJ] {10.1086/505938}, \href
  {2006AJ....132.1221H} {132, 1221}

\bibitem[\protect\citeauthoryear{{Jones}, {R{\"o}pke}, {Pakmor}, {Seitenzahl},
  {Ohlmann}  \& {Edelmann}}{{Jones} et~al.}{2016}]{jonesetal16-1}
{Jones} S.,  {R{\"o}pke} F.~K.,  {Pakmor} R.,  {Seitenzahl} I.~R.,  {Ohlmann}
  S.~T.,   {Edelmann} P.~V.~F.,  2016, \mn@doi [A\&A]
  {10.1051/0004-6361/201628321}, \href {2016A&A...593A..72J} {593, A72}

\bibitem[\protect\citeauthoryear{{Jones} et~al.,}{{Jones}
  et~al.}{2019}]{jonesetal19-1}
{Jones} S.,  et~al., 2019, \mn@doi [A\&A] {10.1051/0004-6361/201834381}, \href
  {2019A&A...622A..74J} {622, A74}

\bibitem[\protect\citeauthoryear{{Kepler} et~al.,}{{Kepler}
  et~al.}{2016}]{kepleretal16-1}
{Kepler} S.~O.,  et~al., 2016, \mn@doi [MNRAS] {10.1093/mnras/stv2526}, \href
  {2016MNRAS.455.3413K} {455, 3413}

\bibitem[\protect\citeauthoryear{{Koester}}{{Koester}}{2009}]{koester09-1}
{Koester} D.,  2009, \mn@doi [A\&A] {10.1051/0004-6361/200811468}, \href
  {2009A&A...498..517K} {498, 517}

\bibitem[\protect\citeauthoryear{{Koester}, {Weidemann}  \&
  {Zeidler}}{{Koester} et~al.}{1982}]{koesteretal82-2}
{Koester} D.,  {Weidemann} V.,   {Zeidler} E.-M.,  1982, A\&A, \href
  {1982A&A...116..147K} {116, 147}

\bibitem[\protect\citeauthoryear{{Kowalski} \& {Saumon}}{{Kowalski} \&
  {Saumon}}{2006}]{kowalski+saumon06-1}
{Kowalski} P.~M.,  {Saumon} D.,  2006, \mn@doi [ApJ Lett.] {10.1086/509723},
  \href {2006ApJ...651L.137K} {651, L137}

\bibitem[\protect\citeauthoryear{{Kramida}, {Ralchenko}  \& {Reader}}{{Kramida}
  et~al.}{2016}]{kramidaetal16-1}
{Kramida} A.,  {Ralchenko} Y.,   {Reader} J.,  2016, in APS Division of Atomic,
  Molecular and Optical Physics Meeting Abstracts. p. Q1.202

\bibitem[\protect\citeauthoryear{{Liebert}}{{Liebert}}{1977}]{liebert77-1}
{Liebert} J.,  1977, \mn@doi [PASP] {10.1086/130077}, \href
  {1977PASP...89...78L} {89, 78}

\bibitem[\protect\citeauthoryear{{Liebert} et~al.,}{{Liebert}
  et~al.}{2003}]{liebertetal03-2}
{Liebert} J.,  et~al., 2003, \mn@doi [AJ] {10.1086/378911}, \href
  {2003AJ....126.2521L} {126, 2521}

\bibitem[\protect\citeauthoryear{{Masci} et~al.,}{{Masci}
  et~al.}{2019}]{mascietal19-1}
{Masci} F.~J.,  et~al., 2019, \mn@doi [PASP] {10.1088/1538-3873/aae8ac}, \href
  {https://ui.adsabs.harvard.edu/abs/2019PASP..131a8003M} {131, 018003}

\bibitem[\protect\citeauthoryear{{Monari} et~al.,}{{Monari}
  et~al.}{2018}]{monarietal18-1}
{Monari} G.,  et~al., 2018, \mn@doi [A\&A] {10.1051/0004-6361/201833748}, \href
  {https://ui.adsabs.harvard.edu/abs/2018A&A...616L...9M} {616, L9}

\bibitem[\protect\citeauthoryear{{Nomoto} \& {Kondo}}{{Nomoto} \&
  {Kondo}}{1991}]{nomoto+condo91-1}
{Nomoto} K.,  {Kondo} Y.,  1991, \mn@doi [ApJ Lett.] {10.1086/185922}, \href
  {https://ui.adsabs.harvard.edu/abs/1991ApJ...367L..19N} {367, L19}

\bibitem[\protect\citeauthoryear{{Nomoto} \& {Leung}}{{Nomoto} \&
  {Leung}}{2017a}]{nomoto+leung17-1}
{Nomoto} K.,  {Leung} S.-C.,  2017a, in {Alsabti} A.~W.,  {Murdin} P.,  eds, 2017, Handbook of Supernovae.
Springer, p.~483

\bibitem[\protect\citeauthoryear{{Nomoto} \& {Leung}}{{Nomoto} \&
  {Leung}}{2017b}]{nomoto+leung17-2}
{Nomoto} K.,  {Leung} S.-C.,  2017b, in {Alsabti} A.~W.,  {Murdin} P.,  eds, 2017, Handbook of Supernovae.
Springer, p.~1275


\bibitem[\protect\citeauthoryear{{Nomoto}, {Thielemann}  \& {Yokoi}}{{Nomoto}
  et~al.}{1984}]{nomotoetal84-1}
{Nomoto} K.,  {Thielemann} F.-K.,   {Yokoi} K.,  1984, \mn@doi [ApJ]
  {10.1086/162639}, \href {1984ApJ...286..644N} {286, 644}

\bibitem[\protect\citeauthoryear{{Pauli}, {Napiwotzki}, {Heber}, {Altmann}  \&
  {Odenkirchen}}{{Pauli} et~al.}{2006}]{paulietal06-1}
{Pauli} E.,  {Napiwotzki} R.,  {Heber} U.,  {Altmann} M.,   {Odenkirchen} M.,
  2006, \mn@doi [A\&A] {10.1051/0004-6361:20052730}, \href
  {2006A&A...447..173P} {447, 173}

\bibitem[\protect\citeauthoryear{{Pelletier}, {Fontaine}, {Wesemael}, {Michaud}
   \& {Wegner}}{{Pelletier} et~al.}{1986}]{pelletieretal86-1}
{Pelletier} C.,  {Fontaine} G.,  {Wesemael} F.,  {Michaud} G.,   {Wegner} G.,
  1986, \mn@doi [ApJ] {10.1086/164410}, \href {1986ApJ...307..242P} {307, 242}

\bibitem[\protect\citeauthoryear{{Piskunov}, {Kupka}, {Ryabchikova}, {Weiss}
  \& {Jeffery}}{{Piskunov} et~al.}{1995}]{piskunovetal95-1}
{Piskunov} N.~E.,  {Kupka} F.,  {Ryabchikova} T.~A.,  {Weiss} W.~W.,
  {Jeffery} C.~S.,  1995, A\&AS, \href {1995A&AS..112..525P} {112, 525}

\bibitem[\protect\citeauthoryear{{Polin}, {Nugent}  \& {Kasen}}{{Polin}
  et~al.}{2019}]{polinetal19-1}
{Polin} A.,  {Nugent} P.,   {Kasen} D.,  2019, \mn@doi [ApJ]
  {10.3847/1538-4357/aafb6a}, \href {2019ApJ...873...84P} {873, 84}

\bibitem[\protect\citeauthoryear{{Raddi}, {Hollands}, {G{\"a}nsicke},
  {Townsley}, {Hermes}, {Gentile Fusillo}  \& {Koester}}{{Raddi}
  et~al.}{2018a}]{raddietal18-2}
{Raddi} R.,  {Hollands} M.~A.,  {G{\"a}nsicke} B.~T.,  {Townsley} D.~M.,
  {Hermes} J.~J.,  {Gentile Fusillo} N.~P.,   {Koester} D.,  2018a, \mn@doi
  [MNRAS] {10.1093/mnrasl/sly103}, \href {2018MNRAS.479L..96R} {479, L96}

\bibitem[\protect\citeauthoryear{{Raddi}, {Hollands}, {Koester},
  {G{\"a}nsicke}, {Gentile Fusillo}, {Hermes}  \& {Townsley}}{{Raddi}
  et~al.}{2018b}]{raddietal18-1}
{Raddi} R.,  {Hollands} M.~A.,  {Koester} D.,  {G{\"a}nsicke} B.~T.,  {Gentile
  Fusillo} N.~P.,  {Hermes} J.~J.,   {Townsley} D.~M.,  2018b, \mn@doi [ApJ]
  {10.3847/1538-4357/aab899}, \href {2018ApJ...858....3R} {858, 3}

\bibitem[\protect\citeauthoryear{{Raddi} et~al.,}{{Raddi}
  et~al.}{2019}]{raddietal19-1}
{Raddi} R.,  et~al., 2019, \mn@doi [MNRAS] {10.1093/mnras/stz1618}, \href
  {https://ui.adsabs.harvard.edu/abs/2019MNRAS.489.1489R} {489, 1489}

\bibitem[\protect\citeauthoryear{{Schatzman}}{{Schatzman}}{1945}]{schatzman45-1}
{Schatzman} E.,  1945, Annales d'Astrophysique, \href
  {https://ui.adsabs.harvard.edu/abs/1945AnAp....8..143S} {8, 143}

\bibitem[\protect\citeauthoryear{{Schlegel}, {Finkbeiner}  \&
  {Davis}}{{Schlegel} et~al.}{1998}]{schlegeletal98-1}
{Schlegel} D.~J.,  {Finkbeiner} D.~P.,   {Davis} M.,  1998, ApJ, \href
  {1998ApJ...500..525S} {500, 525}

\bibitem[\protect\citeauthoryear{{Sch{\"o}nrich}}{{Sch{\"o}nrich}}{2012}]{schoenrich12-1}
{Sch{\"o}nrich} R.,  2012, \mn@doi [MNRAS] {10.1111/j.1365-2966.2012.21631.x},
  \href {https://ui.adsabs.harvard.edu/abs/2012MNRAS.427..274S} {427, 274}

\bibitem[\protect\citeauthoryear{{Sch{\"o}nrich}, {Binney}  \&
  {Dehnen}}{{Sch{\"o}nrich} et~al.}{2010}]{schoenrichetal10-1}
{Sch{\"o}nrich} R.,  {Binney} J.,   {Dehnen} W.,  2010, \mn@doi [MNRAS]
  {10.1111/j.1365-2966.2010.16253.x}, \href
  {https://ui.adsabs.harvard.edu/abs/2010MNRAS.403.1829S} {403, 1829}

\bibitem[\protect\citeauthoryear{{Schwab} \& {Akira Rocha}}{{Schwab} \& {Akira
  Rocha}}{2019}]{schwab+rocha19-1}
{Schwab} J.,  {Akira Rocha} K.,  2019, \mn@doi [ApJ]
  {10.3847/1538-4357/aaffdc}, \href {2019ApJ...872..131S} {872, 131}

\bibitem[\protect\citeauthoryear{{Seitenzahl} \& {Townsley}}{{Seitenzahl} \&
  {Townsley}}{2017}]{seitenzahl+townsley17-1}
{Seitenzahl} I.~R.,  {Townsley} D.~M.,  2017, in \cite{alsabti+murdin17-1},
  p.~1955

\bibitem[\protect\citeauthoryear{{Shen}, {Kasen}, {Miles}  \&
  {Townsley}}{{Shen} et~al.}{2018a}]{shenetal18-2}
{Shen} K.~J.,  {Kasen} D.,  {Miles} B.~J.,   {Townsley} D.~M.,  2018a, \mn@doi
  [ApJ] {10.3847/1538-4357/aaa8de}, \href {2018ApJ...854...52S} {854, 52}

\bibitem[\protect\citeauthoryear{{Shen} et~al.,}{{Shen}
  et~al.}{2018b}]{shenetal18-1}
{Shen} K.~J.,  et~al., 2018b, \mn@doi [ApJ] {10.3847/1538-4357/aad55b}, \href
  {2018ApJ...865...15S} {865, 15}

\bibitem[\protect\citeauthoryear{{Temmink}, {Toonen}, {Zapartas}, {Justham}  \&
  {G{\"a}nsicke}}{{Temmink} et~al.}{2019}]{temminketal19-1}
{Temmink} K.~D.,  {Toonen} S.,  {Zapartas} E.,  {Justham} S.,   {G{\"a}nsicke}
  B.~T.,  2019, A\&A submitted, arXiv:1910.05335

\bibitem[\protect\citeauthoryear{{Tremblay}, {Bergeron}  \&
  {Gianninas}}{{Tremblay} et~al.}{2011}]{tremblayetal11-2}
{Tremblay} P.-E.,  {Bergeron} P.,   {Gianninas} A.,  2011, \mn@doi [ApJ]
  {10.1088/0004-637X/730/2/128}, \href {2011ApJ...730..128T} {730, 128}

\bibitem[\protect\citeauthoryear{{Vennes}, {Nemeth}, {Kawka}, {Thorstensen},
  {Khalack}, {Ferrario}  \& {Alper}}{{Vennes} et~al.}{2017}]{vennesetal17-1}
{Vennes} S.,  {Nemeth} P.,  {Kawka} J.~R.,  {Thorstensen} J.~R.,  {Khalack} V.,
   {Ferrario} L.,   {Alper} E.~H.,  2017, Science, \href {2017Sci...357..680V}
  {357, 680}

\bibitem[\protect\citeauthoryear{{Wegg} \& {Phinney}}{{Wegg} \&
  {Phinney}}{2012}]{wegg+phinney12-1}
{Wegg} C.,  {Phinney} E.~S.,  2012, \mn@doi [MNRAS]
  {10.1111/j.1365-2966.2012.21394.x}, \href {2012MNRAS.426..427W} {426, 427}

\bibitem[\protect\citeauthoryear{{Woosley}, {Taam}  \& {Weaver}}{{Woosley}
  et~al.}{1986}]{woosleyetal86-1}
{Woosley} S.~E.,  {Taam} R.~E.,   {Weaver} T.~A.,  1986, \mn@doi [ApJ]
  {10.1086/163926}, \href {1986ApJ...301..601W} {301, 601}

\bibitem[\protect\citeauthoryear{{York} et~al.,}{{York}
  et~al.}{2000}]{yorketal00-1}
{York} D.~G.,  et~al., 2000, AJ, \href {2000AJ....120.1579Y} {120, 1579}

\makeatother
\end{thebibliography}

\section*{Appendix}

\begin{table*}
\caption{\label{t-lines}
Listed are the transitions used in the abundance analysis (vacuum wavelengths in \AA), additional optical lines that were also used in the analysis of the SDSS discovery spectrum are given in Table\,S2 of \citet{kepleretal16-1}.}
\centering
\begin{tabular}{ll}\hline
Element & Vacuum wavelengths of the main lines [\AA] \\ \hline
H  & optical  \\
He & optical  \\ 
\Ion{C}{i}   & 1277.245 1277.282 1277.513 1277.550 1277.723 1431.596 1432.105 1432.529 1459.031 1463.336 1560.309 1560.682 1560.708  \\
             & 1561.340 1561.438 1656.266 1656.928 1657.008 1657.379 1657.907 1658.121 \\
\Ion{C}{ii}  & 1323.862 1323.906 1323.951 1323.995 1334.532 1335.663 1335.708 \\
\Ion{C}{iii} & 1174.933 1175.263 1175.590 1175.711 1175.987 1176.370 \\
\Ion{N}{i} & 1243.179 1243.306 1243.310 1492.625 1492.820 1494.675 \\
\Ion{O}{i} & optical, 1152.150 1172.500 1172.780 1302.168 1304.860 1306.030 \\
\Ion{O}{ii} & optical, 1502.838 \\
Ne & optical \\
Na &  \\
\Ion{Mg}{ii} & optical, 1365.544 1367.257 1367.708 1369.423 1476.000 1478.004 1480.879 1482.890 1734.852 1737.613 1737.628 1750.664 1753.474 \\
\Ion{Al}{ii} & 1539.833 1625.628 1670.787 1719.442 1721.244 1721.271 1724.949 1724.982 \\
\Ion{Al}{iii} & 1379.670 1384.132 1605.766 1611.814 1611.873 1854.716 1862.790 \\
\Ion{Si}{ii} & optical, 1190.416 1193.290 1194.500 1197.394 1260.422 1264.738 1265.002 1346.884 1348.543 1350.072 1350.516 1350.656 1352.635 1353.721 \\
             & 1508.732 1509.092 1512.064 1513.563 1526.707 1533.431 1710.836 1711.299 1711.304 \\
\Ion{Si}{iii} & 1206.500 1206.555 1207.517 \\
\Ion{P}{ii} & 1159.086 1284.329 1289.569 1294.648 1494.967 1496.439 1506.442 1542.304 1799.060 1799.875 1800.224 1876.777 1879.606 \\
\Ion{P}{iii} & 1334.813 \\
\Ion{S}{i} & 1253.325 1425.030 1425.188 1433.278 1433.309 1807.311 \\
\Ion{S}{ii} & 1115.129 1115.331 1115.605 1115.710 1116.187 1166.291 1166.90 1167.512 1168.150 1124.395 1124.986 1131.059 1131.657 1253.811 \\
\Ion{Ca}{ii} & 1838.008 1840.061 \\
\Ion{Sc}{ii} & 1239.952 1240.415 1240.656 1240.810 1241.166 1241.283 \\
\Ion{Ti}{iii} & 1420.034 1420.439 1421.641 1421.755 1422.409 1455.19 1455.733 \\
\Ion{V}{ii} & 1582.019 1582.344 1582.544 1582.607 1582.855 1634.987 1635.864 1635.866 1636.024 1643.024 1643.056 1643.426 1643.437 \\
\Ion{Cr}{iii} & 1701.478 1701.548 \\
\Ion{Mn}{iii} & 1283.580 1284.064 \\
\Ion{Fe}{ii} & 1558.541 1558.692 1559.085 1563.790 1566.822 1569.675 1570.244 1574.038 1574.922 1578.495 1580.629 \\
\Ion{Fe}{iii} & 1550.154 1550.193 1550.459 1550.862 1551.089 1551.365 1551.392 1552.065 1552.681 1552.936 \\
\Ion{Ni}{ii} & 1164.279 1164.575 1168.041 1381.286 1454.840 1454.842 \\
\Ion{Co}{ii} & 1466.203 1574.545 1576.796 1577.045 \\
\Ion{Cu}{ii} & 1358.773 \\
\Ion{Sr}{ii} & 4078.861 4216.707 \\
\hline
\end{tabular}
\end{table*}

\bsp

\label{lastpage}

\end{document}